\documentclass[fleqn,usenatbib]{mnras}

\usepackage{newtxtext,newtxmath}

\usepackage[T1]{fontenc}
\usepackage{ae,aecompl}

\usepackage{graphicx}	
\usepackage{amsmath}	
\usepackage{amssymb}	
\usepackage[utf8]{inputenc}
\usepackage{float}
\usepackage{ctable}

\title[Slingshot prominences in cool stars]{Prominence formation and ejection in cool stars}

\author[Villarreal D'Angelo et al.]{
Carolina Villarreal D'Angelo $^{1}$\thanks{E-mail: csvd@st-andrews.ac.uk}, Moira Jardine $^{1}$ \& Victor See $^{2}$
\\
$^{1}$SUPA, School of Physics and Astronomy, University of St Andrews, North Haugh, KY16 9SS, UK\\
$^{2}$Department of Physics and Astronomy, University of Exeter, Physics Building, Stocker Road, Exeter, EX4 4QL, UK
}

\date{Accepted 2017 December 19. Received 2017 December 9; in original form 2017 September 1}

\pubyear{2017}

\begin{document}
\label{firstpage}
\pagerange{\pageref{firstpage}--\pageref{lastpage}}
\maketitle

\begin{abstract}
The observational signatures of prominences have been detected in single and binary G and K type stars for many years now, but recently this has been extended to the M dwarf regime.
Prominences carry away both mass and angular momentum when they are ejected and the impact of this mass on any orbiting planets may be important for the evolution of exoplanetary atmospheres.
By means of the classification used in the massive star community, that involves knowledge of two parameters (the co-rotation and Alfv\'en radii, $r_K$ and $r_A$), we have determined which cool stars could support prominences. From a model of mechanical support, we have determined that the prominence mass $m_p/M_\star=(E_M/E_G)(r_\star/r_K)^2 F$ where $E_MB_\star^2r_\star^3$ and $E_G = GM_\star^2/r_\star$ are magnetic and gravitational energies and $F$ is a geometric factor. Our calculated masses and ejection frequencies (typically $10^{16}-10^{17}$g and 0.4 d, respectively) are consistent with observations and are sufficient to ensure that an exoplanet orbiting in the habitable zone of an M dwarf could suffer frequent impacts.
\end{abstract}




\section{Introduction}
Prominences are clouds of cool, mainly neutral gas that are magnetically supported within stellar coronae. On the Sun they are detected mainly in the Balmer series, especially in H $\alpha$, either in absorption on the solar disk, or in emission on the solar limb. Stellar ``slingshot'' prominences are also detected as a transient features in the H $\alpha$ line. The first example was the rapidly rotating K0 star AB Dor \citep{collier1989I,collier1989II,collier1990}, but since then, prominences have been found in fast rotators \citep{Jeffries1993,Byrne1996,Barnes2000}, T Tauri stars \citep{ Eibe1998, Donati2000, Dunstone2006}, binary systems \citep{Steeghs1996, Watson2007, Parsons2011} and recently in M dwarfs \citep{Vida2016}. The intriguing dips seen in K2 observations of many M dwarfs may also be due to a prominence-like phenomenon, where dust may be entrained with the gas \citep{Stauffer2017}.

The accumulation of mass along closed magnetic field structures and its support against centrifugal ejection has been modelled in several different ways. \cite{Ferreira2000} and \cite{Jardine2001} found the existence of stable mechanical equilibria for different types of magnetic configurations. At these points, the component of the effective gravity along the field line is zero and yields a potential minimum. This approach explains the existence of prominences within the closed magnetic field of the stellar corona. The large distances of many prominences from the stellar rotation axis (for example between 3 and 5 stellar radii for AB Dor \citep{collier1989I}), however, requires very extended coronae on these stars. More recently, \cite{Jardine2005} showed that the support of this type of prominences is possible above the cusps of helmet streamers. This allows prominence to exist well beyond the stellar corona, trapped within magnetic loops embedded in the stellar wind.

The issue of plasma accumulation in closed magnetic field structures has also been studied in the massive star community. Rotationally modulated emission in the Balmer lines has been explained, by means of stellar magnetic field confinement, in the massive star $\sigma$ Ori E \citep{Landstreet1978}. These stars have radiatively-driven winds that are powered by the stellar luminosity. The solar-like stars, by comparison, have thermally (or centrifugally) driven winds that are ultimately powered by the stellar magnetic field.
Massive stars have been classified as those with ``centrifugally-supported magnetospheres'' that can  support a stable accumulation of mass in the closed magnetosphere and those with ``dynamical magnetospheres'' where this stable support is not possible \citep{Petit2013}.

In this work, we apply the classification used for massive stars to a sample of 47 low-mass stars whose surface magnetic fields have been mapped. For the prominence-bearing stars, we calculate, the mass, mass loss rate, lifetime and angular momentum loss of the possible prominences.
\section{Input parameters}
Our stellar sample is presented in Table \ref{tab:1}.
Updated values of the surface averaged field strengths of both the axisymmetric component and the non-axisymmetric component of the magnetic dipole are taken from \cite{See2017}. We extract only the dipole term from the full spherical harmonic expansion of the field to ensure consistency within the sample, since the number of harmonics that can be fitted meaningfully depends on the rotation rate of the star. It is also the most relevant mode since it decays most slowly with height above the surface and so will dominate at the heights at which prominences are supported. We use mass loss rates from \cite{See2017}.
These are calculated using the PFFS method of \cite{Altschuler1969} to determine the 3D coronal magnetic field geometry and the relation of \cite{Arge2000} to determine the stellar wind velocity from the expansion of the magnetic field lines. The stellar wind density is obtained by scaling the solar wind density at 1 AU in same manner as \cite{Jardine2008} and \cite{See2015a}. The mass loss rate then follows by integrating over a spherical surface.
We calculate the equatorial Alfvén radius from a Weber-Davis model \citep{Weber1967} using the numerical code from \cite{Johnstone2017}, based on the Versatile Advection Code (VAC) \citep{toth1996} and coronal temperatures calculated using the relation $T_{\rm cor}= 0.11~F_{\rm x}^{0.26}$ from \cite{Johnstone2015}.

\ctable[
	caption= {Physical properties of the star sample taken from \protect\cite{See2015b,See2017} and \protect\cite{Vidotto2014} (based on the BCool and Toupies surveys\protect\tmark). $L_x$ values are taken from \protect\cite{Vidotto2014}, except for the case of V439 And that was taken from \protect\cite{Barbera1993}.},
    label= {tab:1},
    doinside=\normalsize,
    star
]{llllllllll}{
	\tnote{BCool and Toupies surveys published in: Petit (in preparation)  \protect\cite{Boro2015, doNascimiento2014, Donati2003, Donati2008, Fares2009, Fares2010, Fares2012, Fares2013, Folsom2016, Morin2008b,Morin2008a,Morin2010,Jeffers2014,Petit2008} and \protect\cite{Waite2011}.}
}{
	\toprule
    \toprule
	Name & Mass & Radius & $\Omega$& $\langle B_\mathrm{dip} \rangle $&$\dot{M}[\times10^{-12}] $& Log($L_x$)& $T_{cor}$& $r_K$& $r_A$\\
         & $[M_{\odot}]$ & $[r_{\odot}]$ & $[\Omega_{\odot}]$ &[G]& [$M_{\odot}/yr$]  &[erg s$^{-1}$]& [MK] &$[r_*]$&$[r_*]$\\
    \midrule
    \textbf{Solar-like stars} \\
    HD 3651 &	0.88&	0.88&	0.63&	2.90&	0.15&	27.23&	1.68& 	56.6&	10.72\\
    HD 10476&	0.82&	0.82&	1.70&	1.62&	0.08&	27.15&	1.66& 	30.5&	8.21\\
		$\kappa$ Ceti&		1.03&	0.95&	11.37&	10.9&	0.66&	28.79&	4.12&   19.8& 14.3	\\
		$\epsilon$ Eri&	    0.86&	0.77&	9.97&	11.73&	0.40&	28.32&	3.46&	24.6&14.0	\\
		HD 39587&	1.03&	1.05&	5.67&	5.37&	0.77&	28.99&	4.40&	11.5&	7.3\\
		HD 56124&	1.03&	1.01&	1.51&	1.94&	0.13&	29.44&	5.88&	28.9&	5.7\\
		HD 72905&	1.00&	1.00&	5.45&	6.69&	0.60&	28.97&	4.46&	12.3&	9.4\\
		HD 73350&	1.04&	0.98&	2.21&	4.94&	0.45&	28.76&	3.98&	23.2&	8.5\\
		HD 75332&	1.21&	1.24&	5.67&	5.14&	0.61&	29.56&	5.68&   10.3&	8.2\\
		HD 76151&	1.24&	0.98&	1.33&	2.71&	0.20&	28.34&	3.09&	34.5&	8.1\\
		HD 78366&	1.34&	1.03&	2.39&	10.44&	0.82&	28.94&	4.32&	22.8&	13.0\\
		HD 101501&	0.85&	0.90&	1.55&	7.61&	0.37&	28.22&	3.01&	30.0&	13.5\\
		$\xi$ Boo A&    0.85&	0.84&	4.86&	14.07&	0.8&	28.86&	4.58&	15.0&	13.6\\
		$\xi$ Boo B&	0.72&	1.07&	2.64&	9.34&	0.98&	27.97&	2.37&	16.7&	12.7\\
        18 Sco&		0.98&	1.02&	1.20&	0.78&	0.07&	26.8&	1.20&	32.8&	7.9\\
		HD 166435&	1.04&	0.99&	8.01&	8.64&	0.64&	29.5&	6.16&	9.7&	10.2\\
		HD 175726&	1.06&	1.06&   6.98&	4.21&	0.55&	29.1&	4.68&	10.0&	6.7\\
		HD 190771&	0.96&	0.98&	3.10&	6.24&	0.39&	29.13&	4.96&	18.0&	10.1\\
		61 Cyg A&	0.66&	0.62&	0.80&	2.52&	0.06&	28.22&	3.65&	62.2&	7.9\\
		HN Peg&	  	1.085&	1.04&	5.92&	8.87&	0.82&	29&		4.45&	11.5&	10.9\\
		\midrule
        \textbf{Young suns}\\
        AB Dor&		1.0&	1.0&	54.47&	105.10&	7.97&	30.06&	8.57&	2.6&	24.6\\
        HII 739&	1.08&	1.03&	10.09&	7.44&	0.74&	29.33&	5.45&	8.1&	8.9\\
		HIP 12545&	0.58&	0.57&	5.67&	73.94&	2.04&	30.29&	13.18&	17.5&	20.4\\
		V439 And&	0.95&	0.92&	4.39&	9.51& 	0.62&	29.06$^{a}$&	4.92&	15.2&	11.4\\
		\midrule
        \textbf{Hot Jupiter hosts}\\
        $\tau$ Boo&	1.34&	1.42&	9.08&	1.12&	0.25&	28.94&	3.65&	6.8&	4.5\\
		HD 46375&	0.97&	0.86&	0.65&	1.96&	0.09&	27.45&	1.94&	58.5&	9.1\\
		HD 73256&	1.05&	0.89&	1.95&	3.54&	0.22&	28.53&	3.64&	27.9&	8.4\\
        HD 102195&	0.87&	0.82&	2.21&	6.62&	0.25&	28.46&	3.65&	26.1&	12.4\\
		HD 130322&	0.79&	0.83&	1.04&	1.82&	0.08&	27.62&	2.19&	41.2&	8.1\\
		HD 179949&	1.21&	1.19&	3.58&	1.60&	0.18&	28.61&	3.29&	14.6&	6.0\\
		HD 189733&	0.82&	0.76&	2.18&	6.52&	0.33&	28.26&	3.37&	27.9&	10.5\\
		\midrule
        \textbf{M Dwarf stars}\\
		CE Boo&		0.48&	0.43&	1.85&	98.66&	1.55&	28.4&	4.92&	46.0&	31.7\\
		DS Leo&		0.58&	0.52&	1.95&	32.82&	0.53&	28.3&	4.20&	39.2&	23.7\\
		GJ 182&		0.75&	0.82&	6.33&	72.57&	3.59&	29.6&	7.22&	12.3&	25.7\\
		GJ 49&		0.57&	0.51&	1.46&	15.66&	0.27&	28&		3.54&	48.0&	17.2\\
		AD Leo&		0.42&	0.38&	12.38& 	166.10&	3.12&	28.73&	6.39&	14.0&	30.5\\
		DT Vir&		0.59&	0.53&	9.39&	59.15&	1.28&	28.92&	6.03&	13.5&	24.6\\
		EQ Peg A&	0.39&	0.35&	24.76&	365.63&	2.52&	28.83&	7.08&	9.4&	61.9\\
		EQ Peg B&	0.25&	0.25&	68.09&	379.10&	1.75&	28.19&	5.75&	5.8&	54.0\\
		EV Lac&		0.32&	0.3&	6.19&	433.30&	3.25&	28.37&	5.83&	25.8&	60.1\\
		DX Cnc&		0.10&	0.11&	54.47&	63.72&	0.05&	27.61&	6.23&	11.2&	26.8\\
		GJ 1156&	0.14&	0.16&	54.47&	60.53&	0.13&	27.69&	5.38&	8.6&	24.1\\
		GJ 1245B&	0.12&	0.14&	38.91&	79.96&	0.12&	27.35&	4.70&	11.7&	30.4\\
		OT Ser&		0.55&	0.49&	8.01&	66.25&	1.67&	28.8&	5.84&	15.9&	22.7\\
		V374 Peg&	0.28&	0.28&	54.47&	539.61&	2.63&	28.36&	6.01&	6.2&	67.6\\
		WX Uma&		0.10&	0.12&	34.05&	1190.64&1.34&	27.57&	5.81&	14.0&	97.4\\
		YZ Cmi&		0.32&	0.29&	9.73&	526.83&	4.12&	28.33&	5.79&	19.7&	62.7\\
    \bottomrule
    \bottomrule
}

\section{Magnetic confinement-rotation diagram}
\begin{figure}
\includegraphics[width=0.5\textwidth]{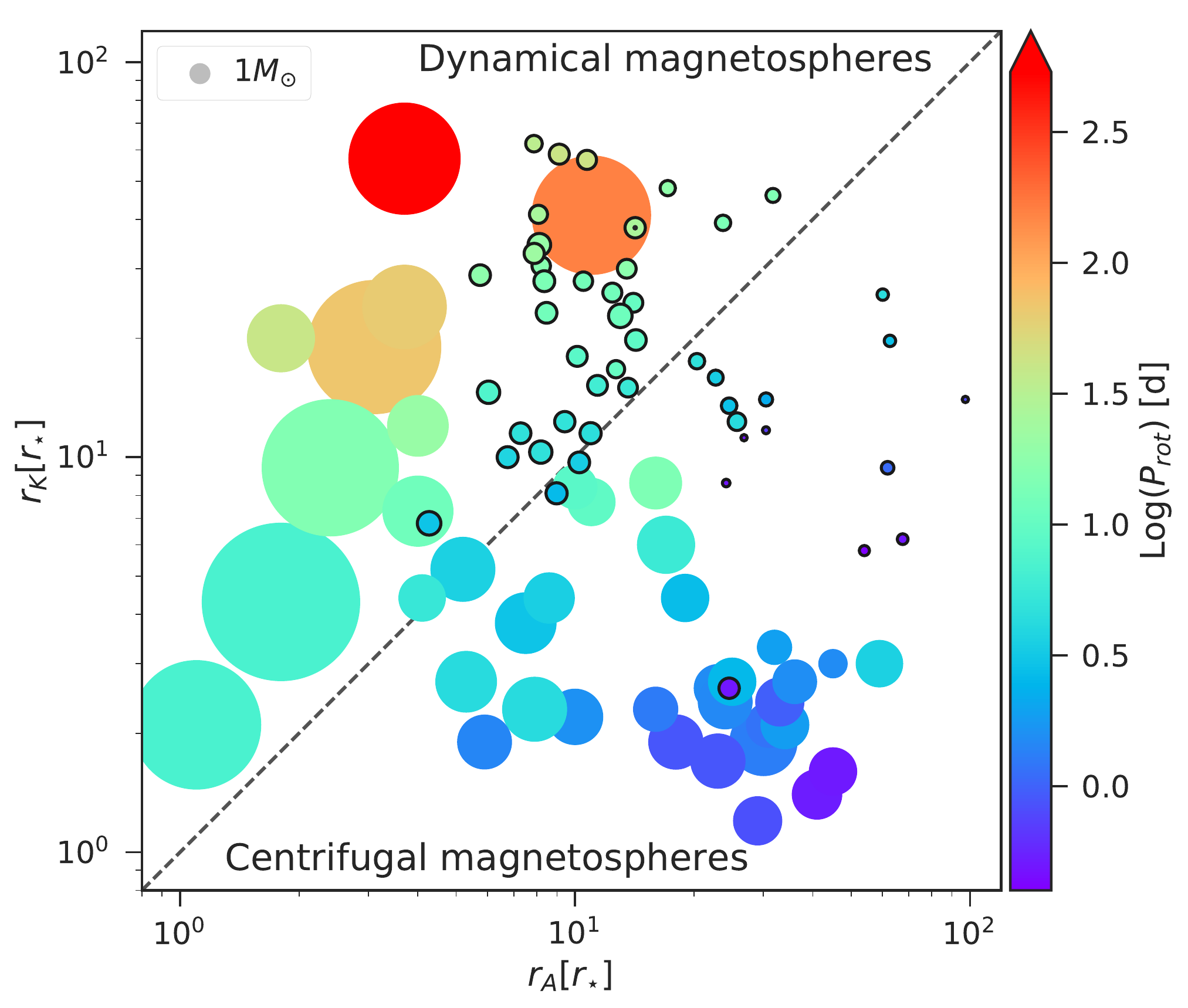}
\caption{Magnetic confinement-rotation diagram: log-log plot of $r_K/r_*$ versus $r_A/r_*$ for the stellar sample in Table \ref{tab:1} (symbols with black border) and the sample of massive stars from the work of Petit et al. (2013). The size of the symbol denotes the stellar mass, while its colour denotes the stellar rotation period. The dashed line indicates $r_K=r_A$. }
\label{fig:1}
\end{figure}

Following \cite{Petit2013} we classify stars by their co-rotation radius ($r_K=~ \sqrt[3]{GM_*/\Omega^2}  $) and their Alfvén radius ($r_A$) where the wind speed equals the local Alfv\'en speed such that $u_{\rm wind} = B/\sqrt{4\pi\rho}$. We show in Fig. \ref{fig:1} both the high-mass stars from \cite{Petit2013} and our sample of low-mass stars.
The Sun is included as reference denoted by the symbol with a black filled circle in the middle.
The stars with ``dynamical magnetospheres'' lie above the dashed line in Fig. \ref{fig:1}. For these stars, $r_K>r_A$ and so any closed magnetic field lines lie below the co-rotation radius. Even if material does accumulate on these field lines, it cannot be supported and so will fall back towards the star on a dynamical time-scale \citep{Sundqvist2012}. In our sample, all solar-like stars and hot Jupiter hosts lie, together with the Sun, in this regime.
Stars with ``centrifugal magnetospheres'' that may be able to support material lie below the dashed line and so have $r_K<r_A$.
Surprisingly, young suns together with M dwarfs in our sample can be found on both sides of the dashed line.

We note that the high- and low-mass stars tend to occupy different parts of this parameter plane.
While the high-mass stars with their dense winds may have smaller co-rotation and Alfv\'en radii, it is the low-mass stars with their lower-density winds that are able to exhibit large co-rotation and Alfv\'en radii.
%
%
Both the high-mass and low-mass stars lie on either side of the $r_A=r_K$ line. However, it is the rapidly-rotating stars within each sample that tend to lie in the centrifugal confinement regime and so may be expected to trap material in their coronae.

\section{Slingshot prominence characteristics}
To estimate the properties of slingshot prominences for the stars in the centrifugal regime, we assume a rigidly-rotating dipolar magnetic field aligned with the stellar rotation axis. Mass accumulates at stable equilibrium points, at a rate determined by the stellar wind, until equilibrium is lost and the material is ejected.

Both \cite{Ferreira2000} and \cite{Townsend2005a} demonstrate that these stable equilibrium points exist in the equatorial plane at a minimum radius of 0.87 $r_K$. They correspond to the minimum of an effective potential calculated along a magnetic field line, and represent probable locations for plasma accumulation. The limit to the mass that can be supported can be estimated by equating the magnetic tension to the effective gravitational force. This gives a maximum density:
\begin{equation}
\rho_b(r) \sim \frac{B^2(r)}{4\pi R_c}\frac{1}{[\Omega^2r - \frac{GM_{\star}}{r^2}]}
\label{eq:1}
\end{equation}
where $R_c$ is the local radius of curvature.
We assume that $R_c$  does not change during the prominence formation, i.e. that the formation time is slow compared to the dynamical time on which the prominence mass is released when equilibrium is lost.
Considering that the prominence has a volume given by its area in the equatorial plane $dA_p = r d\phi dr$
and its height $h_p$, which we take equal to $R_c$ (appropriate for a prominence temperature of $10^5$ K), the mass within a $dV$ is:
\begin{equation}
m_p(r)=\rho_b dA_p h_p.
\end{equation}
The total mass of the prominence is therefore given by integrating over its radial and azimuthal extent to obtain:
\begin{equation}
\frac{m_p}{M_\star}=\frac{E_M}{E_G}\left(\frac{r_\star}{r_K}\right)^2 F,
\label{prom_mass}
\end{equation}
where the magnetic and gravitational energies are $E_M = B_\star^2r_\star^3$ and $E_G = GM_\star^2/r_\star$. F is a geometric factor that accounts for the integration in r and $\phi$.
In terms of $\bar{r}=r/r_K$, F is given by
\begin{equation}
F = \frac{\Delta \phi}{4\pi}\left[
    \frac{1}{\bar{r}^2} -
    \frac{1}{3}\ln\left(  \frac{1-\bar{r}^3}{(1-\bar{r})^2}\right) +
    \frac{2}{\sqrt{3}} \tan^{-1}\left(\frac{2\bar{r}+1}{\sqrt{3}} \right)
\right]
^{\bar{r}_{\rm max}}
_{\bar{r}_{\rm min}}.
\end{equation}

The mass ejection rate from the prominence can be calculated from the mass loss rate of the stellar wind per unit area, multiplied by the footpoint area of the prominence-bearing loop (the factor of 2 accounts for the two foot-points at the star). The relation between the areas is given by the conservation of magnetic flux $dA_p=B_{\star} dA_{\star}/B_p$, where $B_p$ is the magnetic field at the prominence ($B \sim (r_{\star}/r)^3 B_{\star}$). We therefore obtain:
\begin{equation}
\dot{m}_p= \frac{\dot{M}_{\star}}{4\pi r_{\star}^2}2d A_{\star}.
\end{equation}
From this last parameter it is possible to calculate the maximum angular momentum loss, $\dot{J}_p=\dot{m}_p \Omega r^2$, evaluated at the prominence position and their lifetime, $t_p=m_p/\dot{m}_p$.

\begin{figure}
\includegraphics[width = .5\textwidth]{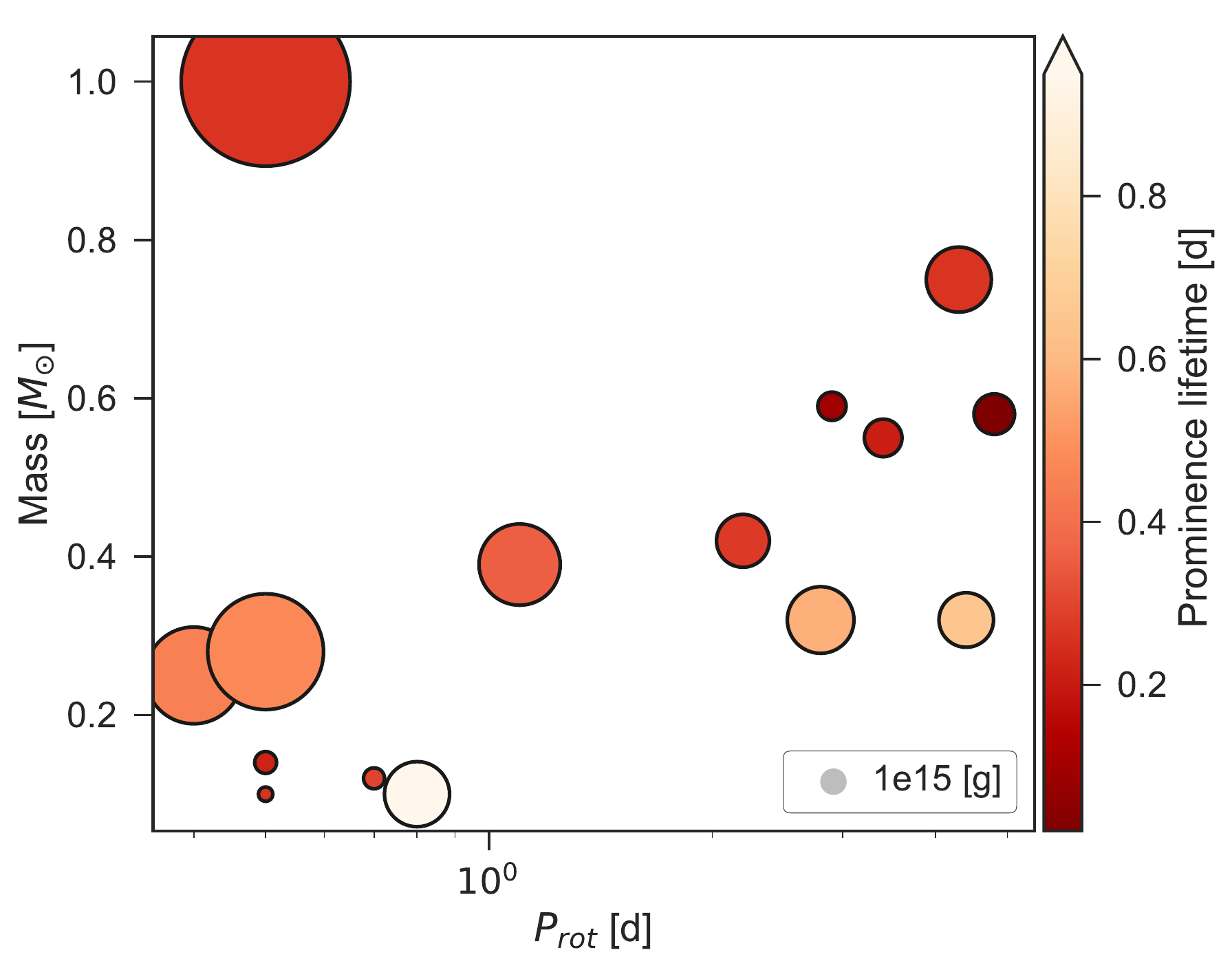}
\caption{Prominence masses (shown by symbol size) and lifetimes (shown by colour) for all the stars in the centrifugal regime presented in Table \ref{tab:2}. The insert shows the symbol size of a typical large solar prominence of $10^{15}$ [g].}
\label{fig:2}
\end{figure}
\section{Results and Discussion}

The prominence characteristics were calculated for those stars in our sample that lie in the centrifugal regime, guided by typical values for  the AB Dor prominences. The integrations were made between $(1.2-1.4)$ $r_K$, corresponding to a radial extent of $(3-5)$ r$_\star$ and an azimuthal extent of $\Delta \phi =40^\circ$ that corresponds to a prominence area covering 20$\%$ of the stellar disc. The results are shown in Table \ref{tab:2} and plotted in Fig. \ref{fig:2}, where colour denotes prominence lifetimes and symbol size denotes the prominence masses scaled to the value of a typical large solar prominence.
We find that the most massive prominences correspond to the fast rotators in our sample: AB Dor, V374 Peg, EQ Peg A and EQ Peg B. The values found for AB Dor are consistent with the ones derived in the work of \cite{collier1990} which reports typical masses between $2-6 \times 10^{17}$ g and typical projected areas of 15-20 $\%$ of the stellar disc. This is several orders of magnitude larger than masses of large solar prominences. Minimum masses of 10$^{16}$g derived from observations of V374 Peg by \cite{Vida2016} and $8 \times 10^{17}$g for AD Leo \citep{Houdebine1990} are also consistent with our values.

Prominence lifetimes depend not only on the prominence mass, but also on the rate at which that mass can be supplied by the stellar wind. Our calculated values show a significant spread, but are of the order of hours, consistent with observations. The rate at which these mass ejections may impact on orbiting planets depends not only on the lifetimes of individual prominences, but also on their number, their latitude of ejection and their rate of expansion, factors which are governed by the geometry of the stellar magnetic field. \citet{Khodachenko2007} estimate that exoplanets orbiting in the habitable zones of M dwarfs would suffer continuous impacts from stellar ejecta if the stellar ejection rate exceeded 36 per day. The M dwarfs in our sample show on average typical lifetimes for prominences of $0.4$ days. If these prominences are continually formed and ejected, this lifetime corresponds to an ejection rate of a minimum of 2 per day for each prominence site. If each prominence has an angular extent of 40$^\circ$ in longitude, this gives an equatorial ejection rate for the star of 18 per day.

The continuous ejection of prominences also comprises a contribution to the mass loss rate from the star. Prominence ejection opens up field lines that would otherwise be closed and allows the closed-field regions of the star to contribute to the overall stellar mass loss. This mass loss may have a different character to the ambient wind that is carried on open field lines, in a similar way to the differences between the fast and slow solar wind components.
For AB Dor, the mass loss carried by each prominence is about 16 $\%$ of the value obtained if the entire stellar surface contributed to the stellar wind.
We note that these may be overestimates, since we are not considering any filamentary structure within the prominence. In future, allowing the prominence material to deform the magnetic field may provide better estimates of prominence mass.
The angular momentum loss through prominence ejection calculated for AB Dor is less than $1\%$ of the angular momentum loss in the wind and remains small for the rest of the sample.
\begin{table}
	\centering
    \setlength\tabcolsep{2.5pt}
	\caption{Physical characteristics of prominences.}
	\label{tab:2}
	\begin{tabular}{lcrccc}
		\hline
        \hline
        Name & $m_p$ & $t_p$ & $\dot{m}_p$  & $\dot{J_p}$  & $ \frac{E_M}{E_G}$\\
             &   [g] &   [d] & [$10^{-12}$ $M_\odot$/yr] & [$10^{32}$ erg]  & [$10^{-13}$ ]  \\
        \hline
        \textbf{Young Suns} \\
        AB Dor	  & 1.9e18 & 0.3  & 1.3e0 & 4.5e0 & 6.9 \\
        HII 12545 & 6.2e15 & 0.02 & 5.6e-2 & 2.9e-1 & 1.8 \\
        \hline
        \textbf{M Dwarf stars}\\
        GJ 182   & 4.1e16 & 0.3   & 2.9e-2 & 1.7e-1 & 4.4 \\
        AD Leo   & 1.8e16 & 0.3   & 1.2e-2 & 3.8e-2 & 4.5 \\
        DT Vir   & 1.5e15 & 0.1   & 2.8e-3 & 1.2e-2 & 0.2 \\
        EQ Peg A & 9.9e17 & 0.4   & 5.1e-2 & 1.2e-1 & 12.0 \\
        EQ Peg B & 2.0e17 & 0.4   & 8.2e-2 & 1.0e-1 & 14.0 \\
        EV Lac   & 2.0e16 & 0.6   & 5.7e-3 & 1.9e-2 & 22.0 \\
        DX Cnc   & 1.0e14 & 0.3   & 7.1e-5 & 5.4e-5 & 0.07 \\
        GJ 1156  & 5.3e14 & 0.2   & 4.3e-4 & 4.1e-4 & 0.1 \\
        GJ 1245B & 4.4e14 & 0.3   & 2.7e-4 & 2.6e-4 & 0.3 \\
        OT Ser   & 4.6e15 & 0.2   & 4.0e-3 & 1.8e-2 & 1.1 \\
        V374 Peg & 4.1e17 & 0.5   & 1.6e-1 & 2.4e-1 & 30.0 \\
        WX Uma   & 4.1e16 & 0.9   & 7.9e-3 & 7.0e-3 & 43.0 \\
        YZ Cmi   & 4.5e16 & 0.6   & 1.4e-2 & 4.2e-2 & 29.0 \\
        \hline
        \hline
   \end{tabular}
\end{table}

\section{Conclusion and summary}

Guided by studies of magnetospheric confinement in massive stars, we have classified a sample of magnetically-mapped low-mass stars by their ability to confine material in their coronae.
We find that despite the very different nature of their winds, low-mass and high-mass stars share common features in their ability to support material within their coronae.
This classification, in terms of the stellar co-rotation radius and wind Alfv\'en radius, reveals that rapidly rotating cool stars are the most likely to support cool clouds (known as ``slingshot prominences'') in their coronae.

Our model of centrifugal support provides the prominence mass in terms of observationally-accessible quantities: the dipole magnetic field strength, the mass, radius and rotation rate of the star. This prominence mass depends on the ratio of the stellar magnetic to gravitational energy, modified by a dependence on rotation rate. This suggests that M dwarfs, which typically show strong magnetic fields and rapid rotation, should host large prominences.

Not all M dwarfs show strong fields, however. The very low-mass stars in the lower left region of Fig. \ref{fig:2} exhibit a bimodal behaviour: either strong, simple fields or weak, complex fields \citep{Morin2011}. Prominence masses in these stars should therefore also show a bimodal distribution.

The dependence of prominence mass on magnetic field strength and rotation rate suggests that as stars lose angular momentum in their winds and therefore spin down, the masses of the prominences they can support should decrease. Stars will evolve out of the regime of ``centrifugally supported magnetospheres'' on a timescale that depends on the spin-down rate of the star.
During this regime, prominences will continually form and be ejected, providing an intermittent addition to the ambient stellar wind.

Both the derived lifetimes and masses of prominences are consistent with observations and suggest that an exoplanet orbiting in the habitable zone of an M dwarf may be exposed to frequent prominence impacts.

\section*{Acknowledgements}
We acknowledge STFC (ST/M001296/1) and H2020 (682393) support and fruitful comments from the referee.




\bibliographystyle{mnras}
\bibliography{villarreal_prominences}


\bsp	
\label{lastpage}
\end{document}